\documentclass[12pt]{article}

\usepackage[letterpaper, margin=1.0in]{geometry}
\usepackage[utf8]{inputenc}
\usepackage[square, authoryear, numbers]{natbib}
\usepackage{rotating,graphicx}
\usepackage{floatflt}
\usepackage{wrapfig}
\usepackage{amsmath}
\usepackage{color}
\usepackage{caption}
\usepackage{overpic}
\usepackage{subcaption}
\usepackage{lineno}
\usepackage{hyperref}
\usepackage{cite}
\usepackage{hyperref}
\usepackage{setspace}
\usepackage{authblk}
\usepackage{algorithm}
\usepackage{algpseudocode}
\usepackage{setspace}

\newcommand{\MW}{\mathrm{M}_\mathrm{W}}

\title{Kinematic Earthquake Sequences on Geometrically Complex Faults}
\author[1,*]{Brendan J. Meade}
\affil[*]{Corresponding author: meade@fas.harvard.edu}
\date{}

\begin{document}
\maketitle

\abstract{Computational earthquake sequence models provide generative estimates of the time, location, and size of synthetic seismic events that can be compared with observed earthquake histories and assessed as rupture forecasts. Here we describe a three-dimensional probabilistic earthquake sequence model that produces slip event time series constrained across geometrically complex non-planar fault systems.  This model is kinematic in nature, integrating the time evolution of geometric moment accumulation and release with empirical earthquake scaling laws. The temporal probability of event occurrence is determined from the time history of geometric moment integrated with short-term Omori-style rate decay following each earthquake achieving long-term time-averaged moment balance. Similarly, the net geometric moment monotonically controls the probability of event localization, and seismic events release geometric moment with spatially heterogenous slip on three-dimensional non-planar fault surfaces.  We use this model to generate a synthetic earthquake sequence on the Nankai subduction zone over a 1,250-year-long interval, including 700+ $\MW{=}5.5{-}8.5$ coseismic events, with decadal-to-centennial scale quiescent intervals quasi-periodic great earthquake clusters followed by aftershock sequences.}



\pagebreak
\section{Introduction}
Earthquake sequence scenarios include estimates of potential event locations, magnitudes, and occurrence times that may be compared to past and future seismicity.  Approaches to sequence generation include the mechanical modeling of time-evolving fault slip and statistical models of seismicity rates.  Mechanics-based earthquake rupture and sequence models predict future fault activity through the direct time integration of the equations that describe friction and stress evolution on fault surfaces \citep[e.g.,][]{rice1996slip, kaneko2008spectral, erickson2020community}.  The governing equations for this class of models are generally linear elasticity to describe stress transfer from one part of a fault to another and some form of rate-and-state friction to describe the time evolution of slip along a discretized representation of the fault surface.  Statistical earthquake occurrence models adopt a less mechanical approach to the space-time evolution of seismicity and instead generate earthquake sequences based on empirically/theoretically motivated functional forms for the probability of event size, location, and time, with earthquakes typically represented as a point source \citep{vere1970stochastic, ogata2011significant} and potentially including time-based representations for strain accumulation and release \citep{rikitake1974probability, matthews2002brownian, salditch2020earthquake, neely2023more}.

Here we introduce a time-dependent stochastic kinematically informed earthquake sequence (SKIES) model developed for the purpose of generating time series of finite-sized seismic events on geometrically complex three-dimensional fault surfaces.  Representing the behavior of individual non-planar fault surfaces across multiple earthquake cycles is a first step towards developing earthquake cycle models that span geometrically complex fault systems \citep[e.g.,][]{plesch2007community, basili2008database}.  The kinematic approach described here avoids explicit stress-based interactions and so does not require volumetric meshing common to finite-element approaches \citep{lo2002finite} nor methods to deal with the hypersingular integrals found in boundary element approaches to elasticity \citep{klockner2013quadrature}. In addition to not seeking to represent elastic fault interactions, it is further not a goal of this model to represent microscale fault physics (rate-state friction, pore fluids, etc.).  

Instead, we develop a mesoscale kinematic representation of time-dependent fault system activity that tracks geometric moment rather than stress and friction. Geometric moment serves as a measure of motion on a fault surface in response to strain energy accumulation and release. Treating strain energy as a primary control on earthquake cycle activity is consistent with the concept that rupture occurrence may be weakly dependent on frictional fault properties and static stress and that ``The only requirement for earthquake rupture propagation is the ability of a frictional system to develop and sustain sufficient stored elastic energy, or ``overstress'', prior to rupture nucleation''\citep{reches_fineberg}.  In part, our approach is inspired by a perspective developed early in the history of thermodynamics, ``In order to consider in the most general way the principle of the production of motion by heat, it must be considered independently of any mechanism or any particular agent. It is necessary to establish principles applicable not only to steam engines but to all imaginable heat engines \textellipsis'' \citep{carnot1824reflections, thurston1897}. Adapting this perspective to the earthquake cycle problem, we describe a model where the production of fault slip is independent of any mechanism or any particular agent so as to establish principles applicable not only to elasticity and rate state friction but to all imaginable rheologies while maintaining a long-term balance of moment accumulation and release rates. Towards this end, the SKIES model enables the probabilistic generation of multiple earthquake sequence scenarios on three-dimensional fault surfaces consistent with empirically constrained seismic scaling laws and geodetic constraints on spatially variable fault loading rates.

\section{Results}
\subsection{Governing concepts for stochastic kinematically informed earthquake sequences}
\begin{figure}[ht]
    \centerline{\includegraphics[width=1.0\textwidth]{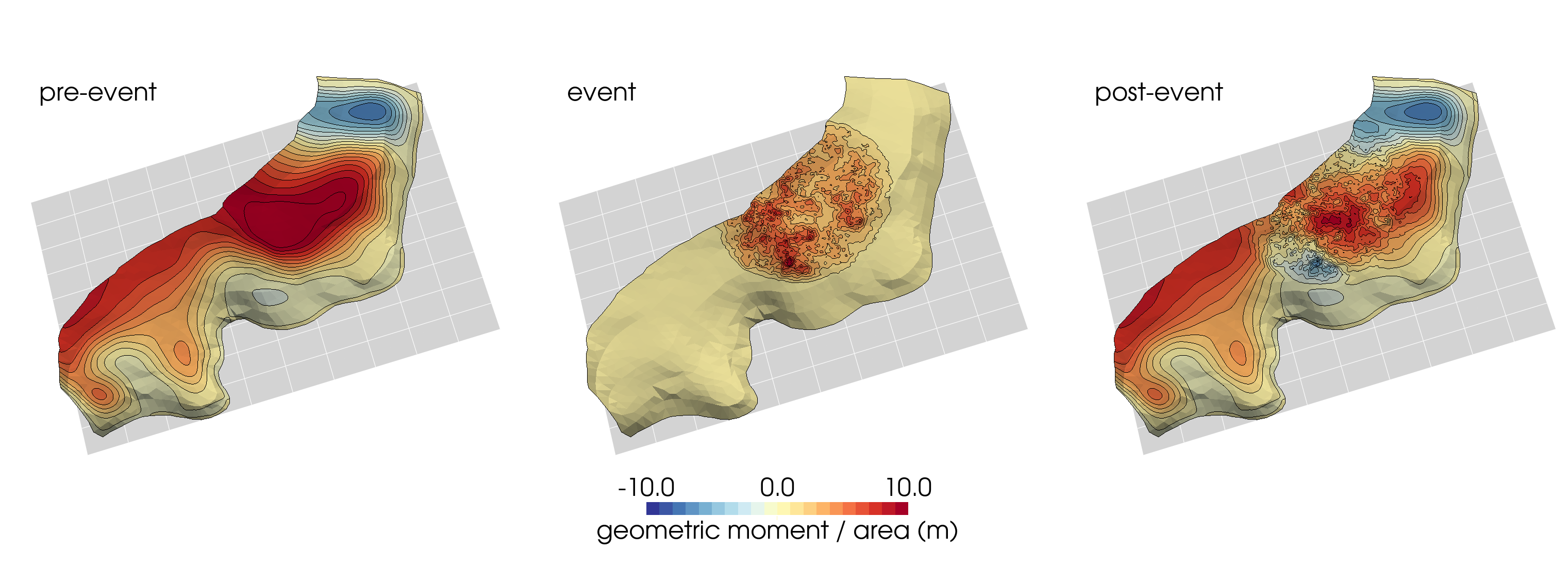}}
    \caption{An example time step in the SKIES model with a coseismic event.  All three panels share the same Nankai subduction zone geometry down to a depth of 50 km, and the view is towards the east-southeast.  The leftmost panel shows the accumulated geometric moment per mesh element area (slip) after 200 years of steady accumulation.  Persistent coseismic sense creep is localized to the southernmost edge of the mesh near Kyushu (blue).  A large event has a hypocentral location near the most strongly coupled region (dark red).  The central panel shows a coseismic slip distribution for a $\MW {=} 8.5$ event.  The coseismic slip is bounded by the mesh geometry extending to and terminating at the trench.  The rightmost panel shows the geometric moment per mesh element area (slip) immediately after the occurrence of the coseismic event.  The geometric moment distribution is unchanged at the extreme northern and southern ends of the subduction zone but decreased in the region of strong pre-event coupling.  Some regions that had positive pre-event $m$ have experienced a coseismic drop significant enough to give a negative post-event $m$.}
    \label{fig:kale_single_step_erosion_figures_nankai}
\end{figure}

\begin{figure}[ht]
    \centerline{\includegraphics[width=0.70\textwidth]{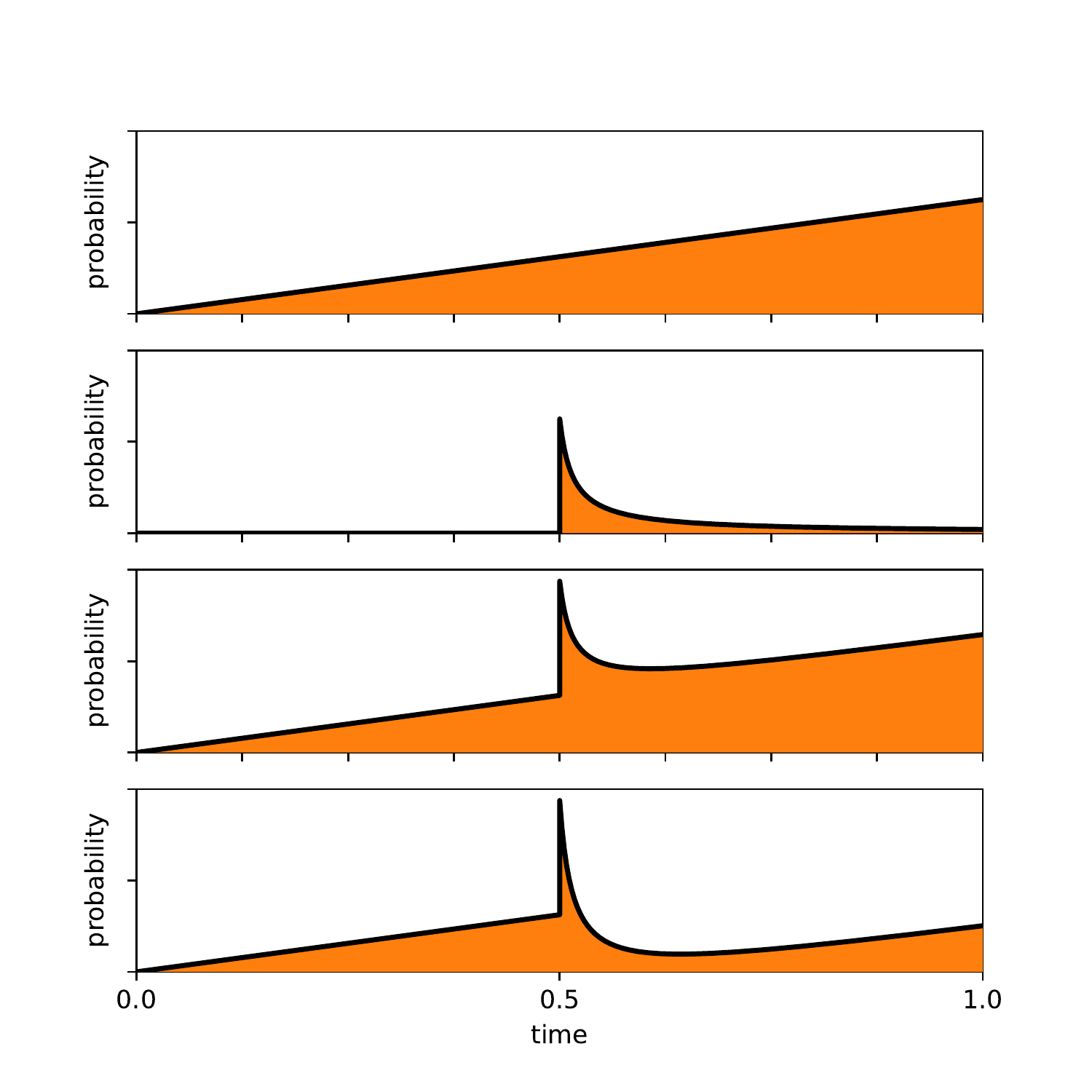}}
    \caption{Schematic diagram of earthquake probability time series. Top panel: long-term tectonic loading.  Second from top panel: Omori-style short-term aftershock excitation following a mainshock at $t=0.5$.  Second from bottom panel: long-term tectonic loading combined with short-term aftershock excitation following a mainshock at $t=0.5$.  Bottom panel: long-term tectonic loading with short-term aftershock excitation following a mainshock at $t=0.5$ and an intermediate-term kinematic decrease associated with the geometric moment release associated with the mainshock at $t=0.5$.  This latter case ensures that the probabilities remain bounded over long time intervals.}
    \label{fig:cartoon_time_series}
\end{figure}

The central goal in the generation of stochastic kinematically informed earthquake sequences (SKIES) is to produce history-dependent sequences of synthetic earthquakes, with spatially variable slip distributions, across a range of magnitudes on non-planar three-dimensional fault surfaces.  The SKIES algorithm is constructed based on the following ten principles. 1) We assume that earthquakes have a tendency to occur more frequently and perhaps be larger in regions with greater loading rates.  Specifically, we consider the loading rate to be the geometric moment accumulation rate, $\mathbf{m}^\mathrm{a}(\mathbf{x}, t)$ that accumulates as a function of slip deficit rates, $\mathbf{v}_\mathrm{sd}$ often constrained from models of nominally interseismic geodetic observations of surface motions. 2) We assume that earthquakes are more likely to be localized in regions characterized by relatively high rates of geometric moment accumulation \citep{moreno20102010, loveless2011spatial}. This direct effect is modulated by the historically contingent and spatially variable geometric moment release due to coseismic events, $\mathbf{m}^\mathrm{r}(\mathbf{x}, t)$.  The net geometric moment, $\mathbf{m}^\mathrm{a} - \mathbf{m}^\mathrm{r}$, serves as the foundation for a probabilistic representation of earthquake centroid locations (figure \ref{fig:kale_single_step_erosion_figures_nankai}). 3) We assume that geometric moment may be positive or negative at any time but remains bounded and balanced (figure \ref{fig:cartoon_time_series}), in an average sense, over long time intervals \citep{segall1986slip, liu20142013, stevens2016millenary}. 4) In the era following coseismic events, there may be a period of enhanced seismicity rate that decays with time following each earthquake \citep[e.g.,][]{omori1895after, utsu1961statistical}. 5) While the Omori effect may generate a period of aftershock-style increased seismicity that decays with time, there is some empirical evidence that large earthquakes themselves may not be particularly clustered in time.  For example, the centennial sequence of large earthquakes along the North Anatolian fault exhibit decadal scale inter-event times \citep[e.g.,][]{barka1996slip}. Thus in the SKIES model, the probability of an event increases temporarily following an initial event due to the Omori effect but decreases over intermediate time scales due to the overall decrease in accumulated geometric moment caused by coseismic geometric moment release (figure \ref{fig:cartoon_time_series}). 6) We assume that earthquake magnitudes are randomly drawn from empirically determined, or hypothesized, magnitude-frequency distributions \citep[e.g.,][]{gutenberg_richter_1944, schwartz1984fault, kagan1997seismic}. 7) Earthquakes with large magnitudes tend to rupture larger areas than those with smaller magnitudes, and we assume that this relationship can be represented using empirical scaling relationships \citep[e.g.,][]{wells1994new, allen2017alternative}. 8) Small events may grow in a self-similar fashion to produce ruptures that are approximately circular \citep[e.g.,][]{burridge1973admissible, miyatake1980numerical, madariaga1998modeling}. 9) Ruptures of all sizes are constrained to occur on a prescribed fault geometry that is subdivided into fault elements.  This is consistent with the observation of ribbon-like ruptures on large continental strike-slip faults \citep[e.g.,][]{lasserre2005coseismic}. 10) Event slip is assumed to be spatially heterogeneous, consistent with inferences from geodetically and seismically informed estimates of coseismic slip distributions \citep[e.g.,][]{mai_spatial_2002, barba2022high}.  

\subsection{Geometric moment and the generation of synthetic earthquake sequences}
Geometric moment, $m$, is a kinematic measure of motion on a fault surface and is defined as the product of fault slip, $s$, and fault area, $a$.  For a discretized representation of a fault surface, the geometric moment on a single mesh element, $i$, is $m_i=s_i a_i$ (no implied summation on repeated indices throughout).  Geometric moment can be accumulated in the interval between large seismic events and released by earthquakes.  Over a time interval spanning $t_0-t$ with $j$ earthquakes occurring at $t_j$, the time evolution of geometric moment on a single fault element in the two-stage (interseismic and coseismic) earthquake cycle case can be written as,

\begin{equation}
    \label{eq:geometric_moment_single_triangle}
    m_i(t) =
     \int_{t_0}^{t} \nu_i a_i v_i^{\mathrm{sd}}(t')dt'- \sum_j^{N_\mathrm{eq}(t_j \leq t)} a_i s_{i,j}(t_j)
\end{equation}

where $\nu_i$ defines the fraction of accumulated geometric moment that is recoverable as coseismic geometric moment release, and $v_i^{\mathrm{sd}}$ is the potentially time-dependent slip deficit velocity, often inferred from the modeling of nominally interseismic geodetic data \citep[e.g.,][]{nishimura2004temporal}.  Over the entire fault surface, the total geometric moment, $\hat{m}(t)$, is the sum of the geometric moment rates from individual fault elements $m_i(t)$.  If the geometric moment is everywhere negative on a fault surface, there is no accumulated geometric moment to be released by earthquakes.  However, if $\hat{m}(t)$ is negative, it could be the case that all fault elements have negative $m_i$ or that the contribution from fault elements with negative $m_i$ exceeds that of fault elements with positive $m_i$.

\subsection{Coseismic event generation}
In the SKIES model, whether a coseismic event occurs at a specific time is determined by a weighted Bernoulli draw of a time probability weight, $\mathcal{B}(1, p^t)$.  The time probability weight, $p^t$, of a single earthquake occurring at time $t$ can be represented as a function of four contributions: 1) a rate that is a function of geometric moment accumulation, $r^\mathrm{a}$, 2) an Omori-style increase in probability following discrete events $r^\mathrm{o}$, 3) a function of geometric moment release  $r^\mathrm{r}$, 4) and an auxiliary term, $\mathcal{A}$, that could represent additional perturbations due to external forcing (e.g., information transferred from earthquakes outside of the meshed region) or potential noise contributions,

\begin{equation}
    p^t = \gamma_a^t \tanh \left( \gamma_d^t \left[r^\mathrm{a} + \sum\nolimits_j^{n(t_j \leq t)} \{ r^\mathrm{o} + r^\mathrm{r} \} + \mathcal{A} \right] \right).
    \label{eq:time_probability_weight}
\end{equation}

The geometric moment accumulation rate term is the sum of the accumulated geometric moment accumulation, $a_i v_i^{\mathrm{sd}}$, over all mesh elements, modulated by a mesh element dependent term $\nu_i$ that reflects a fraction of the geometric moment rate that is not releasable through coseismic events (e.g., lost to plastic deformation), $r^\mathrm{a}(t) = \sum\nolimits_i \int_{t_0}^{t} \nu_i a_i v_i^{\mathrm{sd}}(t')dt'$.  A rate increase of event probability following all events at times, $t_j$, is given by a classical Omori-style response $r^\mathrm{o}_j(t) = \beta_j / [1 + (t - t_j)^{p_j} / \tau_j]$, where $\beta_j$ is the productivity of the $j^\mathrm{th}$ event, $t_j$ is the event time, $p_j$ is a potentially magnitude dependent Omori exponent, and $\tau_j$ is a magnitude dependent relaxation time \citep[e.g.,][]{ouillon_and_sornette_2005}. Note that all events, regardless of magnitude, size, location, and time of occurrence, have an Omori-style probability response, and no event is characterized as a foreshock, mainshock, or aftershock.  The rate change due to moment release can be written as, $r^\mathrm{r} = -\omega' \beta' \left[ \sum\nolimits_i m_i(t_j) \right] ^{\psi'}$, where $\omega'$,  $\beta'$, and $\psi'$ are constants.  This term describes the mechanism by which coseismic geometric moment release may lead to a decrease in intermediate-term (e.g., 10-100 years) event probabilities following the short-term increase in event probability associated with the Omori effect (figure \ref{fig:cartoon_time_series}).  Again, other functional forms may be applicable as well.  Similarly, the use of $\tanh$ as a normalizing function is arbitrary and could be replaced with other viable options or potentially be a functional form that is solved for given some constraining observations.  The case, $\omega' = \beta' = \psi' = 1$, would represent the case where the time probability weight over long time intervals is directly modulated by moment balance constraints. This non-linear relationship between geometric moment accumulation and event probability saturates with geometric moment, and we further bound it with minimum ($10^{-6}$) and maximum ($0.05$) probabilities for the example below.

A coseismic slip event will occur at the current time step, $t_k$,  in a discretized model realization if a weighted Bernoulli draw for event occurrence, $\mathcal{B}(1, p^\mathrm{t}(t_k))$ is equal to one.  To generate a synthetic event, we randomly draw the event magnitude, $\MW$, from a prescribed magnitude-frequency distribution (e.g., Gutenberg-Richter, $10^{a-b\mathrm{M_W}}$) bound between specified minimum and maximum magnitudes.  The event moment geometric moment is calculated from the moment to moment magnitude relationship, $m = \mu^{-1}10^{1.5 \mathrm{M_W} + 9.05}$, \citep{hanks_and_kanamori_1979} with an assumed shear modulus, $\mu$.  The fact that the event area scaling laws are determined from regressions to $\MW$ rather than $m$ is the only reason we introduce an assumed shear modulus.  An approximate event rupture area, $\hat{a}$, can be determined from empirical scaling laws \citep{allen2017alternative}.  We further multiply this area by a scaling factor ${>}1$ (1.2 for the examples here) to create a slightly larger potential rupture area on which we will realize a static spatially heterogenous coseismic slip distribution that tapers to zero slip at its lateral extent.

The location of the current event, $x_i^{\mathrm{h}}$, is determined by selecting a random element from a triangulated mesh surface with a probability of element selection proportional to the $p_i^h(t_k) = \gamma_a^h \tanh ( \gamma_d^h \left[m^\mathrm{a}_i - m^\mathrm{r}_i\right] )$ using a weighted Bernoulli draw $x_i^{\mathrm{h}} = \mathcal{B}(1, p_i^\mathrm{h}(t_k))$.  Again, the choice of a hyperbolic tangent as a normalizing function is arbitrary (as with $p^t$) and could be replaced with alternatives (e.g., a sigmoid function).  With the hypocentral triangle, $x_i^{\mathrm{h}}$, and event area known, we then identify the subset of triangular mesh elements, $l$, surrounding $x_i^{\mathrm{h}}$ that will provide a cumulative area $\hat{a}\approx\sum a_l$.  We select elements $x_l$ with Euclidean distance closest to $x_i^{\mathrm{h}}$ such that the area of a rupture is approximately circular away from the boundaries of the fault mesh.  Slip events are constrained to lie entirely on a prescribed gap-free mesh, and so if the accumulated event area reaches a mesh edge while still less than $\hat{a}$, the next elements selected for the event accumulated area will be selected along the boundary of the mesh where geometrically limited and radially otherwise.  For non-planar surfaces, the effective distance from $x_i^{\mathrm{h}}$ to $x_l$ would be more accurately calculated using the along mesh surface geodesic distance rather than the Euclidean distance used here.

The mean event slip is then given by $\hat{s} = m/\sum a_l$, and we generate spatially heterogeneous static coseismic slip distributions on the subset of identified event elements largely following \citep{leveque_2016} and \citep{Melgar_LeVeque_Dreger_Allen_2016} where the central idea is to randomly weight eigenmodes determined from the Karhunen-Lo\`eve expansion of the distances between mesh element centroids.  For the subset of mesh elements involved in a slip event, we calculate a Gaussian correlation matrix based on the pairwise relative distances between mesh elements involved in a particular rupture, $C_{mn} = \exp({-|\mathbf{x}_m-\mathbf{x}_n|})$ for all $ m, n$ in $l$.  The set of eigenvectors, $\mathbf{S}$, eigenmodes, $\mathbf{V}$, on the slipping elements is determined from the singular value decomposition of $\mathbf{C}$, and we assign each eigenmode a random weight drawn from a Gaussian distribution.  Other distributions are feasible for the generation of stochastic slip distributions \citep[e.g.,][]{mai_spatial_2002, small_melgar}.  An initial distribution of spatially variable fault slip is constructed by combining the $\mathbf{s} = \exp({\sum_p^{n_\mathrm{eigs}} w_p \mathbf{V}_{p} \sqrt{S_p} })$, where $p \leq l$.  This produces an event slip distribution that has heterogeneous slip, but that slip may abruptly change from some finite value to zero at the edge of the rupture mesh elements on which the event is constrained to lie.  We opt to have slip taper smoothly towards zero at the edges of the rupture extent by multiplying the slip distribution by a radial sigmoid function, $\mathrm{Sigmoid}(|\mathbf{x}^h-\mathbf{x}_l|, \, \rho)$, with characteristic length scale, $\rho$.  Other options are viable as well, including slip tapering to zero with increasing depth \citep{leveque_2016}.  The final event slip distribution is determined by rescaling the slip magnitudes on all patches to match the scalar target geometric moment $s_l \gets s_l m / \sum a_l s_l$.

\section{A 1,000+ Year Long Synthetic Earthquake Sequence on the Nankai Subduction Zone}
Here we use the SKIES algorithm to generate a synthetic earthquake sequence on a representation of the Nankai subduction zone, which sits at the northwestern boundary of the Philippine Sea plate and subducts beneath parts of the Kyushu, Shikoku, and Honshu islands of the Japanese archipelago.  We represent the Nankai subduction zone as a contiguous mesh with 1,804 triangular elements with a total area of  ${\sim} 205,094$ km$^2$ \citep{hirose2008three}. Spatially variable slip deficit rates, $v^\mathrm{sd}(\mathbf{x})$, on this interface are estimated using a three-dimensional block model with slip deficit rates estimated using a three-dimensional block model \citep{loveless2010geodetic, loveless2015kinematic} constrained by GPS velocities \citep{sagiya2004decade} and with the assumption that slip deficit rates taper to zero on deepest, southernmost, and northernmost edges of the subduction zone mesh.  These geodetically constrained slip deficit rates (averaging 25 mm/yr, figure \ref{fig:kale_single_step_erosion_figures_nankai}) are smoothly interpolated onto a higher resolution mesh with 121,728 mesh elements ranging in area from 0.1 to 3.7 km$^2$ to enable the realization of spatially heterogenous slip for small magnitude events. 

\begin{figure}[ht]
    \centerline{\includegraphics[width=1.0\textwidth]{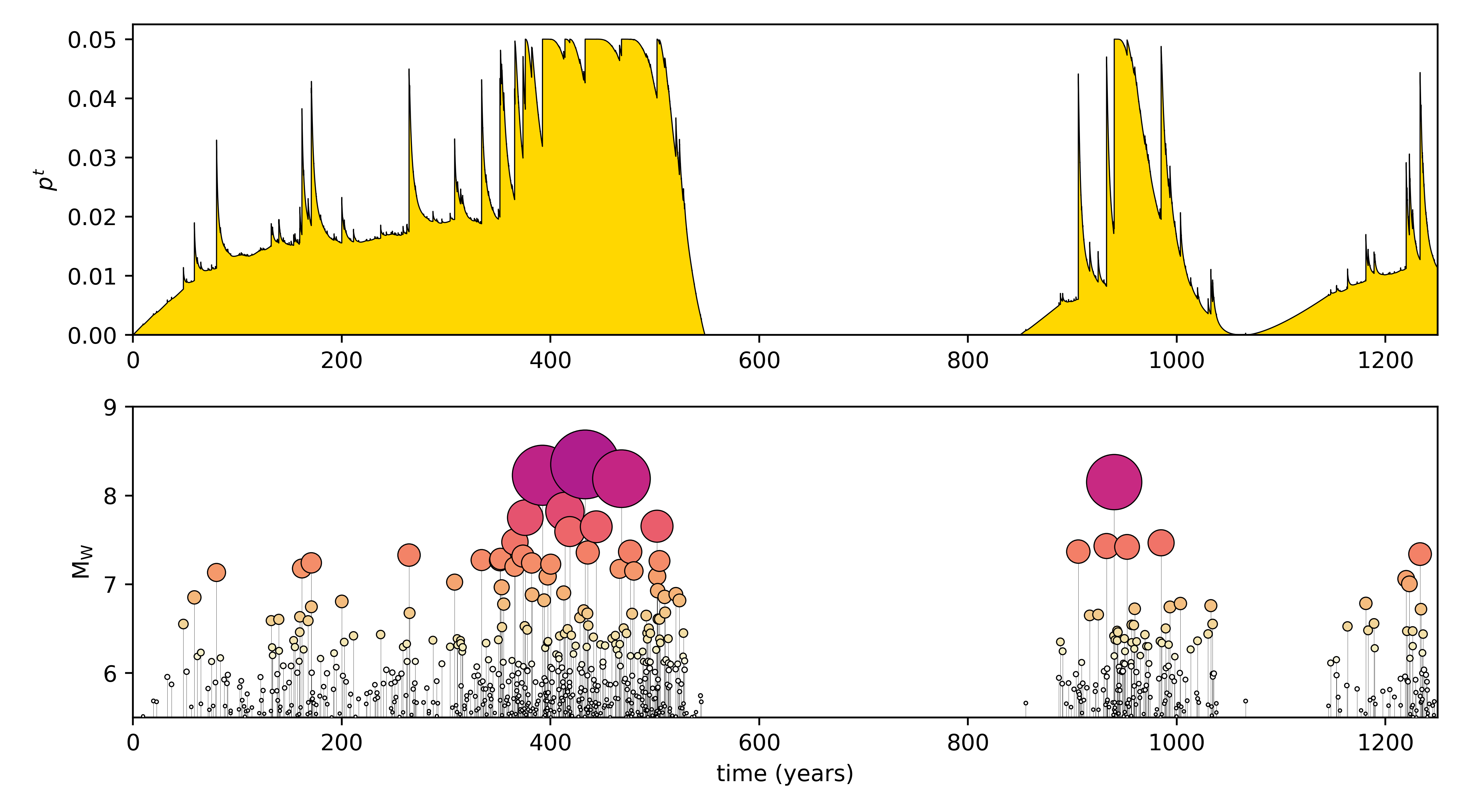}}
    \caption{Time series of event time probability $p^t$ (yellow shaded region in upper panel) and times and magnitudes of coseismic events (lower panel) from a single SKIES model run.  Here the maximum event probability of an event occurring at any time step is limited to 0.05.  The probability of event occurrence increases with time tectonic loading and is enhanced immediately following all earthquakes but effectively decreases over intermediate time scales as the aftershock effect decreases and the overall decrease in the total geometric moment due to the geometric moment release due to large events (equation \ref{eq:time_probability_weight}). The event magnitudes in the lower panel are indicated by the height, the symbol area, and color (darker colors are larger magnitude events).  Large events are followed by short-term aftershock sequences and occasional periods of extended quiescence.  In this realization, there are two periods of relative quiescence a longer 300-year period and a shorter 150-year period.}
    \label{fig:probability_magnitude}
\end{figure}

\begin{figure}[ht]
    \centerline{\includegraphics[width=1.0\textwidth]{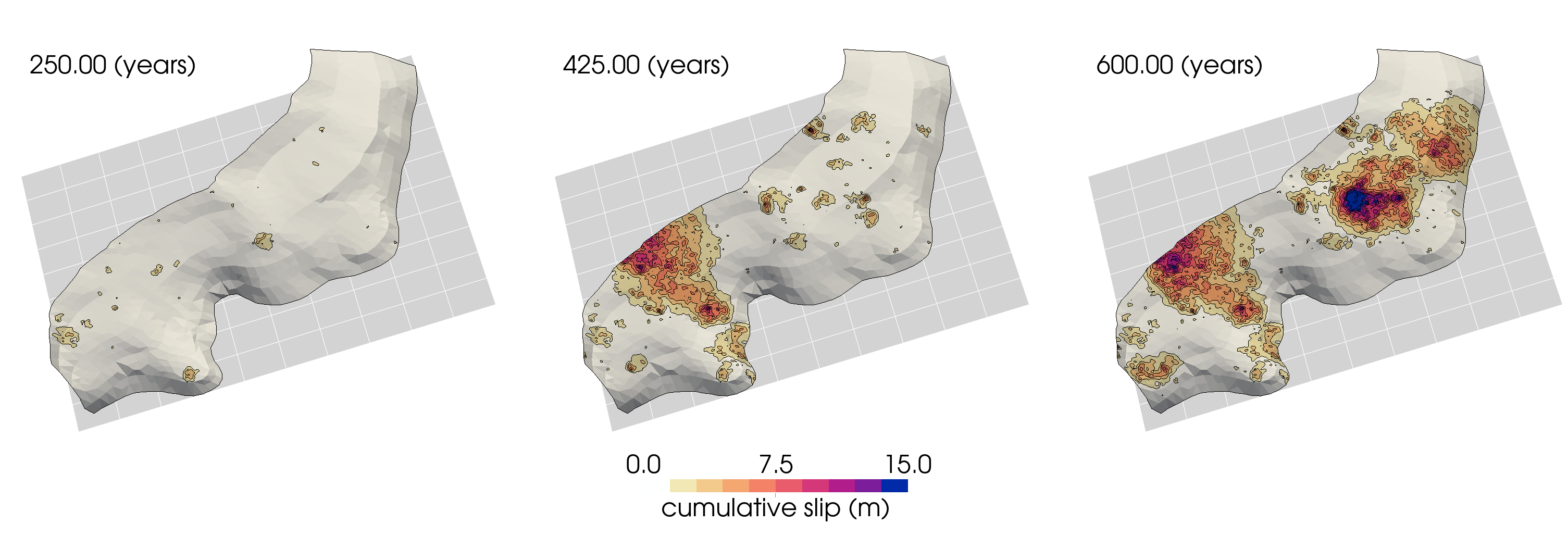}}
    \caption{Cumulative fault slip at three time steps from a SKIES realization of an earthquake cycle sequence on the Nankai subduction zone.  Fault geometry is the same as figure \ref{fig:kale_single_step_erosion_figures_nankai}.  From left to right, the panels show the cumulative fault slip at 250, 425, and 600 years.  Between the 150 and 425 snapshots, a magnitude $\MW{=}8.1$ event ruptures to the top of the trench near the northern edge of the subduction zone, along with ${\sim}150$ smaller events.  A second $\MW{>}8.0$ occurs farther to the south (but still north of the creeping regions) with a rupture extent that is not limited by fault geometry due to the deeper rupture depth}
    \label{fig:kale_single_time_step_cumulative_slip_nankai}
\end{figure}

For the coseismic sequence generation, event moment magnitudes are limited to $\MW{=}5.5-9.0$ and auxiliary forcing is assumed to be negligible $\mathcal{A}=0$.  This synthetic sequence is active for 1,250 years with time steps of ${\sim}9.1$ days generating 723 coseismic events (figure \ref{fig:probability_magnitude}).  The initial ${\sim}50$ years of the time series exhibit an approximately linear increase in event probability.  This is caused by this steady loading, a lack of large coseismic events, and the nearly linear behavior of $\tanh$ for arguments ${<}0.5$.  After 50 years, the $p^t$  increases more slowly, primarily due to the initial occurrence of $\MW {\geq} 6.5$ events that are large enough to decrease the net geometric moment significantly.  During the $t=50{-}350$ epoch $\MW {\geq} 5.5$ events occur every ${\sim}5$ years with no events larger than $\MW {\geq} 7.3$.  After this, a period of enhanced seismicity begins with a series of five $\MW{>}7.1$ events over a 40-year period before a series of three $\MW{>}8$ events with aftershock sequences occurs over the next 130 years (figure \ref{fig:probability_magnitude}).  Note that we limit $p^t$ to a maximum value of 0.05.  Following this epoch of enhanced large-magnitude seismicity, there is a period of extended quiescence lasting 340 years until $\MW {\sim} 6.5$ magnitude seismicity returns at ${\sim}890$ years.  During this quiescent period, $\hat{m}<0$ but $p^t$ is forced to stay slightly positive (${\geq} 10^{-6}$), allowing for a small probability of event occurrence as geometric moment slowly accumulates.  The second active period lasts only 170 years with only one $\MW{>}8$ event (figure \ref{fig:kale_single_time_step_cumulative_slip_nankai}) before a second quiescent period begins at ${\sim}1,040$ years and lasting for ${\sim}100$ years.  This second quiescent period differs from the first in that $\hat{m}$ is positive almost throughout, and yet the random drawing of events with relatively low $p^t$ happens to not yield significant seismic activity. The next active period of large-magnitude seismicity starts near year 1,140 and continues until the end of the model run without the occurrence of a $\MW{\geq}8$ event.  Note that throughout the entire synthetic event sequence, there are no coseismic events on the southernmost part of the Nankai subduction zone (figure \ref{fig:kale_single_time_step_cumulative_slip_nankai}) due to the fact that there is no moment accumulation in this part of the model because of the geodetic inference of slow coseismic sense slip \citep{loveless2010geodetic}.

\begin{figure}[ht]
    \centerline{\includegraphics[width=0.6\textwidth]{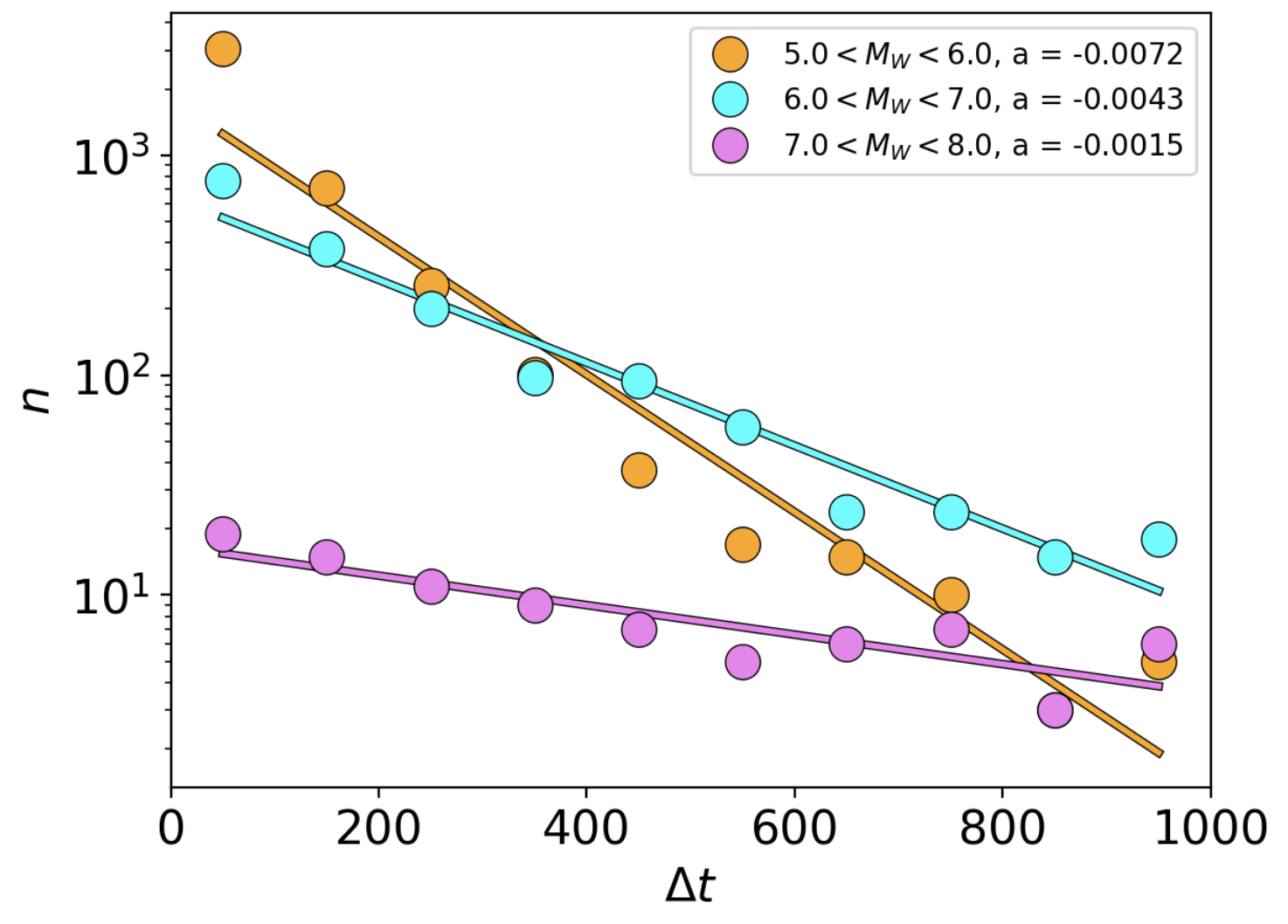}}
    \caption{Interevent recurrence interval to event frequency scaling.  Interevent times, $\Delta t$ (shown as time steps), are calculated as the time between events in sequential time within a magnitude range.  For the three cases considered here ($5.0 {<} \MW {<} 6.0$ orange circles, $6.0 {<} \MW {<} 7.0$ cyan circles, $7.0 {<} \MW {<} 8.0$ magenta circles), we find a greater frequency of short interevent times (100 time steps) as compared with (1,000 time steps) that can be approximated as a power-law: $n\sim \Delta t^{a({\MW})}$.  The power-law exponent exhibits a magnitude dependence decreasing as $\MW$ increases.  This scaling indicates that smaller magnitude events have a greater probability of being clustered in time than larger events.}
    \label{fig:magnitude_recurrence_time}
\end{figure}

For events in a given magnitude range, we determine the time to the next event for events in the same magnitude range over the complete time series and then consider the frequency of these recurrence times (figure \ref{fig:magnitude_recurrence_time}).  For a Nankai model with lower spatial resolution but run for 8,000+ years, we find a range for interevent times at the three magnitude ranges considered ($5.0{<}\MW{<}6.0$, $6.0{<}\MW{<}7.0$, and $7.0{<}\MW{<}8.0$) and that small events tend to occur more closely in time to one another.  The clustering of absolute (not normalized by geometric moment accumulation time scales appropriate for each $\MW$) interevent times decreases with event magnitudes as events in the $7.0{<}\MW{<}8.0$ as these events are more homogeneously distributed in time.  This particular model run did not produce sufficient $\MW{>}8.0$ events to estimate whether great earthquake interevent times homogenize or become localized and well-defined periodicity.

\section{Discussion}
The generation of multi-decade decreases in event probability following large earthquakes is a result of the interplay between geometric moment release rate formulation, $r^\mathrm{r}$, and the Omori-style rate increase, $r^\mathrm{o}$.  In the quasi-static formulation used here, both of these effects occur instantly after every earthquake with  $r^\mathrm{r}$ decreasing the probability of another event and $r^\mathrm{o}$ increasing the probability of another event.  In the absence of an Omori effect, $r^\mathrm{o}=0$, the geometric moment release rate term would mean that there would be no aftershock interval and there would be a drop in event probability compared to the pre-event probability.  This is conceptually similar to the stress shadow \citep{harris_suppression_1998}.  However, with $r^\mathrm{o} {>} 0$ there is a short-term increase in event probability that is larger than the immediate decrease due to $r^\mathrm{r}$.  The aftershock effect decays in time, and if the decay is quick enough, there a period of low probability follows before the slow tectonic loading increases the accumulated moment and $p^t$ (figures \ref{fig:cartoon_time_series}, \ref{fig:probability_magnitude}).  This description provides an overview of the competing rate effects but, in the context of the probabilistic nature of this model, does not limit the suite of possible behaviors.  For example, just because a large earthquake (e.g., $\MW{\geq}8.0$) has occurred, the probabilistic nature of this model means that it is not necessarily the case that another large cannot occur in the near future.  For example, in the Nankai realization described above, we see a sequence of three $\MW{\geq}8.0$ events occurring within 100 years of each other (figure \ref{fig:probability_magnitude}) before a long period of quiescence and is potentially consistent with a supercycle of the clustered complementary rupture type\citep{philibosian2020segmentation}. Longer model runs could reveal whether or not larger ruptures $\MW{>}9.0$ are produced by this model exhibit behavior of the rare multi-asperity ruptures combined with clustered complementary rupture supercycle type. 

The kinematic earthquake sequence model described here provides a means of developing earthquake scenarios consistent with empirical scaling laws, present-day slip deficit rates, and past earthquakes recorded in the geologic record.  Specifically, the comparison of synthetic sequences with geological event chronologies \citep[e.g.,][]{berryman2012major} provides a means of estimating constraints model potential constraints $\omega'$, $\beta'$, $\psi'$ in the geometric moment release rate term, $r^\mathrm{r}$, as well as the minimum and maximum bounds on event probability $p^t$ and event magnitude.  Constraints on these parameters, and their uncertainties, would provide a basis for generating calibrated ensembles of models near the constrained values that produce synthetic earthquake sequences that could be evaluated against the locations, magnitudes, and timing of future seismicity.

Beyond the direct application to seismic event forecasting, the SKIES model can be coupled to a mechanical deformation model to serve as a source of earthquake cycle activity that contributes to the growth of geological structures in response to repeated earthquakes \citep{elliott2016himalayan, dal2021building}.  This offers the possibility of identifying deep histories of possible past earthquakes that are consistent with not only empirical earthquake scaling laws and geometric moment balance but also topography at seismically active plate boundaries.

\pagebreak
\bibliographystyle{agufull08}
\bibliography{references.bib}

\pagebreak
\section*{Acknowledgments}
The authors declare that they have no competing interests.  The authors acknowledge that they received no funding in support for this research.

\section*{Open research}
The SKIES software used for all calculations in the paper is preserved at DOI:10.5281/zenodo.7770999, and developed openly at: \url{https://github.com/brendanjmeade/skies}.

\pagebreak

\pagebreak

\pagebreak

\pagebreak

\pagebreak

\end{document}